\documentclass[structabstract]{aa}
\usepackage{graphicx}
\usepackage{color}
\usepackage{amsmath}
\usepackage{amssymb}
\usepackage{latexsym}
\usepackage{times}
\usepackage[varg]{txfonts}
\begin{document}
   \title{Kelvin-Helmholtz instability on coronal mass ejecta in the \\ lower
   corona}

   \author{I.~Zhelyazkov$^1$, T.~V.~Zaqarashvili$^{2,3}$, and R.~Chandra$^4$}

   \institute{$^1$ Faculty of Physics, Sofia University,
              5 James Bourchier Blvd., 1164 Sofia, Bulgaria\\
              \phantom{$^1$ }\email{izh@phys.uni-sofia.bg}\\
              $^2$ Space Research Institute, Austrian Academy of Sciences,
              Schmiedlstrasse 6, Graz, Austria\\
              $^3$ Abastumani Astrophysical Observatory at Ilia State University,
              3/5 Cholokashvili Avenue,\\
              \phantom{$^2$ }0162 Tbilisi, Georgia \\
              $^4$ Department of Physics, DSB Campus, Kumaun University, Nainital 263\,002, India}

   \date{Received August 12, 2014}


  \abstract
  {} 
   {We model an imaged Kelvin-Helmholtz (KH) instability on a coronal mass ejecta (CME) in the lower corona by investigating conditions under which kink ($m = 1$) and $m = -3$ magnetohydrodynamic (MHD) modes in an uniformly twisted flux tube moving along its axis become unstable.}
   {We employ the dispersion relations of MHD modes derived from the linearised magnetohydrodynamic equations.  We assume real wave numbers and complex angular wave frequencies, namely complex wave phase velocities.  The dispersion relations are solved
   numerically at fixed input parameters (taken from observational data) and various mass flow velocities.}
   {It is shown that the stability of the modes depends upon four parameters, the density contrast between the flux tube and its environment, the ratio of the background magnetic fields in the two media, the twist of the magnetic field lines inside the tube, and the value of the Alfv\'en Mach number (the ratio of the tube velocity to Alfv\'en speed inside the flux tube).
   For a twisted magnetic flux tube at a density contrast of $0.88$, background magnetic field ratio of $1.58$, and a normalised magnetic field twist of $0.2$, the critical speed for the kink ($m = -3$) mode (where $m$ is the azimuthal mode number) is $678$~km\,s$^{-1}$ just as it is observed. The growth rate for this harmonic at KH wavelength of $18.5$~Mm and ejecta width of $4.1$~Mm is equal to $0.037$~s$^{-1}$, in agreement with observations. KH instability of the $m = -3$ mode may also explain why the KH vortices are seen only at the one side of arising CME.}
   {The good agreement between observational and computational data shows that the imaged KH instability on CME can be explained in terms of emerging KH instability of the $m = -3$ MHD mode in twisted magnetic flux tube moving along its axis.}


   \authorrunning{I.~Zhelyazkov et al.}
   \titlerunning{Kelvin-Helmholtz instability on CME}
   \maketitle
%

\section{Introduction}
\label{sec:intro}
Coronal mass ejections (CMEs) are huge clouds of magnetized plasma that erupt from the solar corona into the interplanetary space. They propagate in the heliosphere with velocities ranging from $20$ to $3200$~km\,s$^{-1}$ with an average speed of $489$~km\,s$^{-1}$, based on SOHO/LASCO coronagraph (Brueckner \cite{brueckner1995}) measurements between 1996 and 2003.
CMEs are associated with enormous changes and disturbances in the coronal magnetic field.  The \emph{magnetic breakout}, the \emph{torus instability}, and the \emph{kink instability\/} are the main physical mechanisms that initiate and drive solar eruptions (see, e.g., Aulanier \cite{aulanier2014}, Schmieder et al.\ \cite{schmieder2013}; Chandra et al.\ \cite{chandra2011}; Aulanier et al.\ \cite{aulanier2010}; Forbes et al.\ \cite{forbes2006}; T{\"o}r{\"o}k \& Kliem \cite{torok2005}; Sakurai \cite{sakurai1976}, and references therein).  CMEs are a key aspect of coronal and interplanetary dynamics and are known to be the major contributor to severe space weather at the Earth. Observations from SOHO, TRACE, Wind, ACE, STEREO, and SDO spacecrafts, along with ground-based instruments, have improved our knowledge of the origins and development of CMEs at the Sun (Webb \& Howard \cite{webb2012}).

It is well-known that the solar atmosphere is magnetically structured and in many cases the plasma in so-called magnetic flux tubes is flowing.  Tangential velocity discontinuity at a tube surface due to the stream motion of the tube itself with respect to the environment leads to the Kelvin-Helmholtz (KH) instability which via KH vortices may trigger enhanced magnetohydrodynamic (MHD) turbulence.  While in hydrodynamic flows KH instability can develop for small velocity shears (Drazin \& Reid \cite{drazin1981}), a flow-aligned magnetic field stabilizes plasma flows (Chandrasekhar \cite{chandrasekhar1961}) and the instability occurrence may require larger velocity discontinuities.  For the KH instability in various solar structures of flowing plasmas (chromospheric jets, spicules, soft X-ray jets, solar wind) see, e.g., Zaqarashvili et al.\ (\cite{zaqarashvili2014a}) and references therein.

When CMEs start to rise from the lower corona upwards, then the velocity discontinuity at their boundary may trigger KH instability. Recently, Foullon et al.\ (\cite{foullon2011}) reported the observation of KH vortices at the surface of CME based on unprecedented high-resolution images (less than $150$~Mm above the solar surface in the inner corona) taken with the Atmospheric Imaging Assembly (AIA) (Lemen et al.\ \cite{lemen2012}) on board the Solar Dynamics Observatory (SDO) (Dean Pesnell et al.\ \cite{pesnell2012}). The CME event they studied occurred on 2010 November 3, following a C4.9 GOES class flare (peaking at 12:15:09 UT from active region NOAA 11121, located near the southeast solar limb).  The instability was detected in the highest AIA temperature channel only, centered on the $131$~\AA~EUV bandpass at $11$~MK.  In this temperature range, the ejecta lifting off from the solar surface forms a bubble of enhanced emission against the lower density coronal background (see Fig.~1 in Foullon et al.\ (\cite{foullon2011})).  Along the northern flank of the ejecta, a train of three to four substructures forms a regular pattern in the intensity contrast.  An updated and detailed study of this event was done by Foullon et al.\ (\cite{foullon2013}).  A similar pattern was reported by Ofman \& Thompson (\cite{ofman2011}) -- the authors presented observations of the formation, propagation, and decay of vortex-shaped features in coronal images from the SDO associated with an eruption starting at about 2:30 UT on 2010 April 8.  The series of vortices were formed along the interface between an erupting (dimming) region and the surrounding corona.  They ranged in size from several to $10$~arc sec and traveled along the interface at $6$--$14$~km\,s$^{-1}$.  The features were clearly visible in six out of the seven different EUV wave bands of the AIA. Later, using the AIA on board the SDO, M\"{o}stl et al.\ (\cite{meostl2013}) observed a CME with an embedded filament on 2011 February 24, revealing quasi-periodic vortex-like structures at the northern side of the filament boundary with a wavelength of approximately $14.4$~Mm and a propagation speed of about $310 \pm 20$~km\,s$^{-1}$.  These structures, according to authors, could result from the KH instability occurring on the boundary.

The aim of this paper is to model the imaged KH instability by Foullon et al.\ (\cite{foullon2013}) via investigating
the stability/instability status of tangential velocity discontinuity at the boundary of the ejecta.  The paper is organized as follows.  In Sec.~\ref{sec:obs}, we present a brief description about the observations on 2010 November, 3. In Sec.~\ref{sec:basic} we specify the geometry of the problem, governing equations and the derivation of the wave dispersion relation.  Sec.~\ref{sec:dispers} deals with its numerical solutions for specified, according to observational data, values of the input parameters, as well as the extracting KH instability parameters from computed dispersion curves and growth rates of the unstable $m = 1$ and $m = -3$ harmonics.  The last Sec.~\ref{sec:concl} summarises the results derived in this paper.

\section{Observational description of 2010 November 3 event}
\label{sec:obs}
\begin{figure}[ht]
\centering
  \begin{minipage}{\columnwidth}
  \centering
    \includegraphics[width=7.0cm]{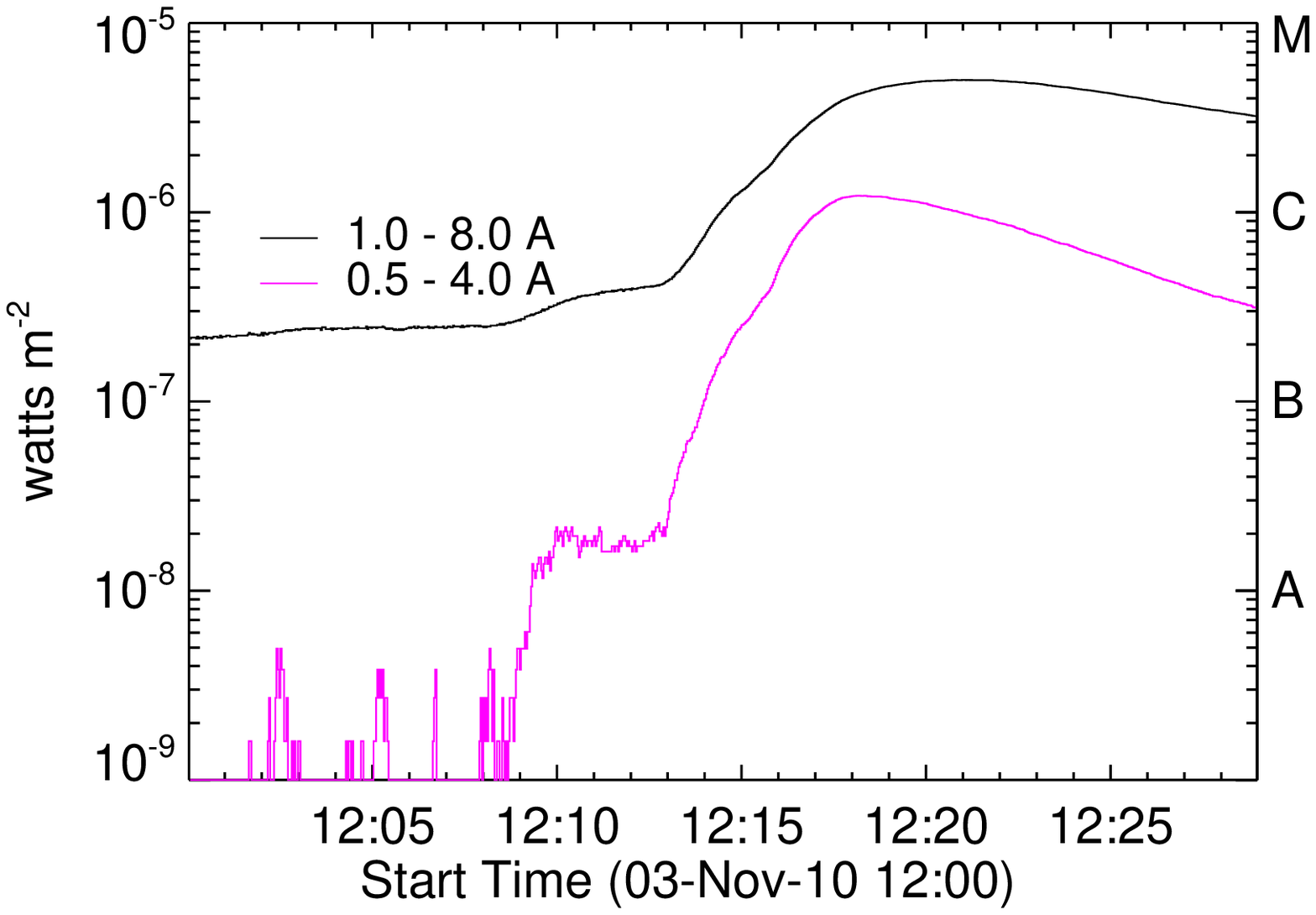} \\
\vspace{0mm}
    \includegraphics[width=7.0cm]{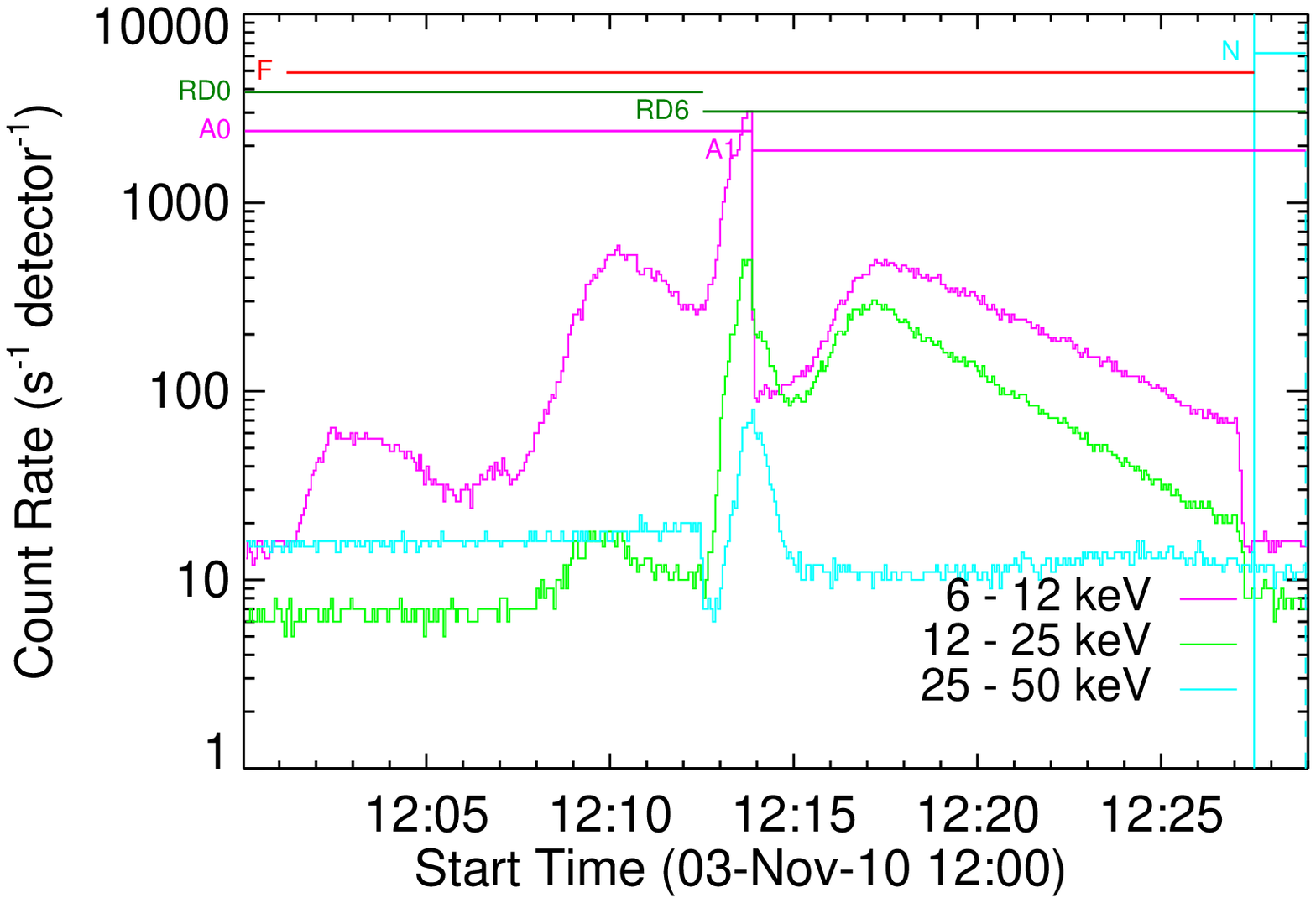}
  \end{minipage}
  \caption{(\emph{Top panel}) GOES time profiles of C4.9 flare in two wavelengths ($1$--$8$ and $0.5$--$4.0$~\AA). (\emph{Bottom panel}) RHESSI time profiles of same flare at different energy bands.}
\label{fig:fig1}
\end{figure}

We present here the observations of GOES C4.9 class flare on 2010 November 3 associated with CME from the NOAA active region 11121. The active region was located at S21E88 on that day on the solar disk.  According to GOES observations the flare was initiated at 12:07 UT, peaked 12:15 UT and ended around 12:30 UT.  The flare/eruption was well observed by SDO/AIA at different wavelengths with high spatial ($0.6$~arcsec) and temporal evolution ($12$~s) as well as by Reuven Ramaty High-Energy Solar Spectroscopic Imager (RHESSI, Lin et al.\ \cite{lin2002}) in X-ray $3$--$6$~keV, $6$--$12$~keV, $12$--$25$~keV, and $25$--$50$~keV energy bands, respectively.  The temporal evolution of flare in X-rays observed by GOES and RHESSI is presented in Fig.~\ref{fig:fig1}.

\begin{figure}[ht]
\centering
  \begin{minipage}{\columnwidth}
  \centering
    \includegraphics[width=9.35cm]{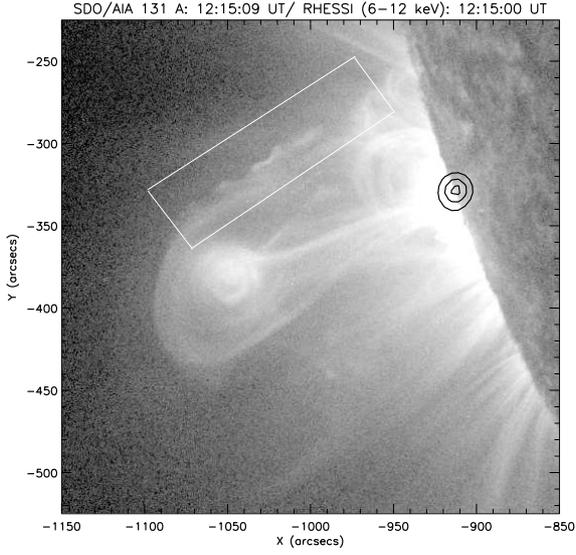} \\
\vspace{-5mm}
    \includegraphics[width=7.0cm]{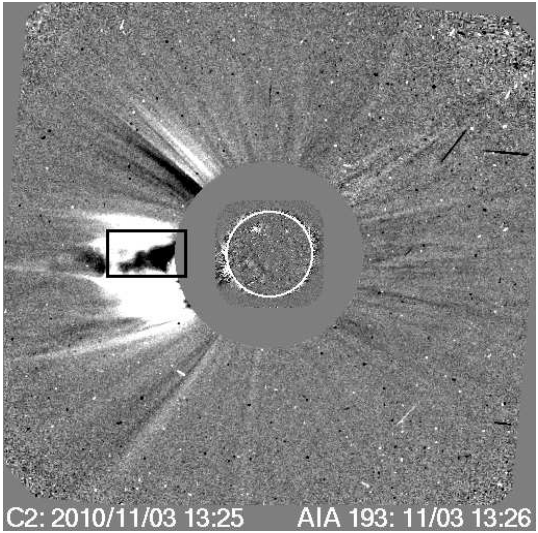}
  \end{minipage}
  \vspace{-10mm}
  \caption{(\emph{Top panel}) SDO/AIA $131$~\AA~image overlaid by RHESSI X-ray contours (contour level: $40\%$, $70\%$, and
  $95\%$ of the peak intensity).  The rectangular white box represents where the KH instability is visible. (\emph{Bottom panel})
  The associated difference image of LASCO CME overlaid on the SDO/AIA image.}
  \vspace*{-1mm}
\label{fig:fig2}
\end{figure}

The SDO/AIA observations on 2010 November 3 at different wavelengths indicate the evidence of KH instability around 12:15 UT at $131$~\AA~(see also Foullon et al.\ \cite{foullon2011} and Foullon et al.\ \cite{foullon2013}). Figure~\ref{fig:fig2} (top panel) shows the SDO/AIA $131$~\AA~at 12:15 UT, where we can see the KH instability features (see the white rectangular box).  The image is overlaid by the RHESSI $6$--$12$~keV contours.  The RHESSI image is constructed from seven collimators (3F to 9F) using the CLEAN algorithm, which yields a spatial resolution of ${\sim}7$~arcsec (Hurford et al.\ \cite{hurford2002}).

LASCO observed a CME at 12:36 UT in C2 field-of-view with speed of $241$~km\,s$^{-1}$ and angular width of  $66^{\circ}$ associated with the flare.  The difference LASCO C2 images overlaid on SDO $193$~\AA~difference image is presented in the bottom panel of Fig.~\ref{fig:fig2}.  In this image, some step-like structure is also visible, which corresponds to KH instability features visible in the the SDO $131$~\AA~wavelengths.  This CME image indicates that we can see the KH instability features up to a ${\sim}3$\,$R_{\odot}$ height.

The study by Foullon et al.\ (\cite{foullon2013}) of the dynamics and origin of the CME on 2010 November 3 by means of the Solar TErrestrial RElations Observatory Behind (STEREO-B) located eastward of SDO\ by $82^{\circ}$ of heliolongitude, and used in conjunction with SDO give some indication of the magnetic field topology and flow pattern.  At the time of the event, Extreme Ultraviolet Imager (EUVI) from STEREO's Sun--Earth Connection Coronal and Heliospheric Investigation (SECCHI) instrument suite (Howard et al.\ \cite{howard2008}) achieved the highest temporal resolution in the $195$~\AA~bandpass: EUVI's images of the active region on the disk were taken every $5$ minutes in this bandpass.  The authors applied the Differential Emission Measure (DEM) techniques on the edge of the ejecta to determine the basic plasma parameters -- they obtained electron temperature of $11.6 \pm 3.8$~MK and electron density $n = (7.1 \pm 1.6)\times 10^8$~cm$^{-3}$, together with a layer width $\Delta L = 4.1 \pm 0.7$~Mm.  Density estimates of the ejecta environment (quiet corona), according to Aschwanden \& Acton (\cite{aschwanden2001}), vary from ($2$ to $1$)$\times10^8$~cm$^{-3}$ between $0.05$ and $0.15$~$R_{\odot}$ ($40$--$100$~Mm), at heights where Foullon et al.\ (\cite{foullon2013}) started to see the KH waves developing.  The final estimation based on a maximum height of $250$~Mm and the highest DEM value on the northern flank of the ejecta yields electron density of $(7.1 \pm 0.8)\times 10^8$~cm$^{-3}$.  The adopted electron temperature in the ambient corona is $T = 4.5 \pm 1.5$~MK.  The other important parameters derived on using the pressure balance equation assuming a benchmark value for the magnetic field $B$ in the environment of $10$~G are summarized in Table~2.  The main features of the imaged KH instability presented in Table~3 include (in their notation) the speed of $131$~\AA~CME leading edge, $V_{\rm LE} = 687$~km\,s$^{-1}$, flow shear on the $131$~\AA~CME flank, $V_1 - V_2 = 680 \pm 92$~km\,s$^{-1}$, KH group velocity, $v_{\rm g} = 429 \pm 8$~km\,s$^{-1}$, KH wavelength, $\lambda = 18.5 \pm 0.5$~Mm, and exponential linear growth rate, $\gamma_{\rm KH} = 0.033 \pm 0.012$~s$^{-1}$.

\section{Geometry, the governing equations, and the wave dispersion relation}
\label{sec:basic}
\begin{figure}[ht]
   \centering
   \includegraphics[height=.25\textheight]{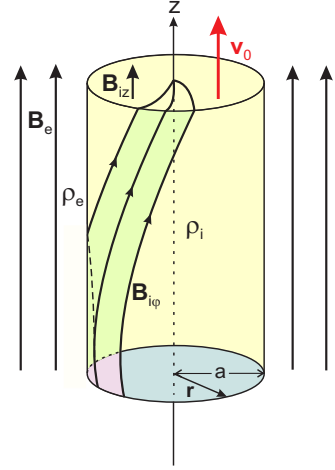}
   \caption{Magnetic field geometry of a coronal mass ejection. \emph{Picture credit to: R.~Erd\'elyi \& V.~Fedun 2010.}}
   \label{fig:fig3}
\end{figure}

As we already said, coronal mass ejections are ejections of magnetized plasma from the solar corona.  Recently, Vourlidas (\cite{vourlidas2014}) presented observational evidence for the existence of magnetic flux ropes within CMEs. The observations detect the formation of the flux rope in the low corona, reveal its sometimes extremely fast evolution and follow it into interplanetary space. The results validate many of the expectations of the CME initiation theories.  The idea for formation of magnetic flux ropes associated with the CMEs is not new.  Manchester et al.\ (\cite{manchester2005}), in studying coronal mass ejection shocks and sheath structures relevant to particle acceleration in the solar wind, by means of a three-dimensional numerical ideal MHD model, drive a CME to erupt by introducing a Gibson-Low magnetic flux rope that is embedded in the helmet
streamer in an initial state of force imbalance. The flux rope rapidly expands and is ejected from the corona with maximum speeds in
excess of $1000$~km\,s$^{-1}$, driving a fast-mode shock from the inner corona to a distance of $1$~AU. Kumar et al.\ (\cite{kumar2012}) presenting multi wavelength observations of helical kink instability as a trigger of a CME which occurred in active region NOAA 11163 on 2011 February 24, state that the high-resolution observations from the SDO/AIA suggest the development of helical kink instability in the erupting prominence, which implies a flux rope structure of the magnetic field (see also Srivastava et al.\ \cite{srivastava2010} for observed kink instability in the solar corona). A brightening starts below the apex of the prominence with its slow rising motion (${\sim}100$~km\,s$^{-1}$) during the activation phase.  In his review of recent studies on coronal dynamics: streamers, coronal mass ejections, and their interactions, Chen (\cite{chen2013}) claims that the energy release mechanism of CMEs can be explained through a flux rope magnetohydrodynamic model. According to him, the rope is formed through a long-term reconnection process driven by the shear, twist, and rotation of magnetic footpoints; moreover, recent SDO studies discovered solar tornadoes providing a natural mechanism of rope formation (see, e.g., Zhang \& Liu \cite{zhang2011}; Li et al.\ \cite{li2012}; Wedemeyer-B\"{o}hm et al.\ \cite{wedemeyer2012}).  Recent \emph{in-situ\/} measurements also suggest that the twisted magnetic tubes could be common in the solar wind (Zaqarashvili et al.\ \cite{zaqarashvili2014b}).  Thus, we are convinced to accept the expanded definition of the CME suggested in Vourlidas et al.\ (\cite{vourlidas2013}): ``A CME is the eruption of a coherent magnetic, twist-carrying coronal structure with angular width of at least $40^{\circ}$ and able to reach beyond $10$~$R_{\odot}$ which occurs on a time scale of a few minutes to several hours.''  Accordingly, the most appropriate model of the CME under consideration is a twisted magnetic flux tube of radius $a$ (${=}\Delta L/2$) and density $\rho_{\rm i}$ embedded in a uniform field environment with density $\rho_{\rm e}$ (see Fig.~\ref{fig:fig3}).  The magnetic field inside the tube is
helicoidal, $\vec{B}_{\rm i} = (0, B_{{\rm i} \varphi}(r), B_{{\rm i} z}(r))$, while outside it is uniform and directed along the $z$-axis, $\vec{B}_{\rm e} = (0, 0, B_{\rm e})$.  The tube is moving along its axis with speed of $\vec{v}_0$ with regards to the
surrounding medium.  The jump of tangential velocity at the tube boundary then triggers the
magnetic KH instability when the jump exceeds a critical value.

In cylindrical equilibrium the magnetic field and plasma pressure satisfy the equilibrium condition in the radial direction
\begin{equation}
\label{eq:equilib}
    \frac{\mathrm{d}}{\mathrm{d}r}\left( p_{\rm i} + \frac{B_{\rm i}^2}{2\mu}
    \right) = -\frac{B_{{\rm i} \varphi}^2}{\mu r}.
\end{equation}
Here, $B_{\rm i}(r) = \left( B_{{\rm i} \varphi}^2 + B_{{\rm i} z}^2 \right)^{1/2} = \left| \vec{B}_{\rm i} \right|$ denotes the strength of the equilibrium magnetic field, and $\mu$ is the magnetic permeability.  We note that in Eq.~(\ref{eq:equilib}) the total (thermal plus magnetic) pressure gradient is balanced by the tension force (the right-hand side of Eq.~(\ref{eq:equilib}))
in the twisted field.  We consider the special case of an equilibrium with uniform twist, i.e., the one for which $B_{{\rm i} \varphi}(r)/r B_{{\rm i} z}(r)$ is a constant.  Thus the background magnetic field is assumed to be
\begin{equation}
\label{eq:magnfield}
    \vec{B}(r) = \left\{ \begin{array}{lc}
                     (0, Ar, B_{{\rm i} z}) & \mbox{for $r \leqslant a$}, \\
                     (0, 0, B_{\rm e}) & \mbox{for $r > a$},
                              \end{array}
                      \right.
\end{equation}
where $A$, $B_{{\rm i} z}$, and $B_{\rm e}$ are constant.  Then the equilibrium condition (\ref{eq:equilib}) gives the equilibrium plasma pressure $p_{\rm i}(r)$ as
\[
    p_{\rm i}(r) = p_0 - \frac{A^2 r^2}{\mu},
\]
where $p_0$ is the plasma pressure at the center of the tube.  We note that the jump in the value of $B_{\varphi}(r)$ across $r = a$ implies a surface current there.

Prior to starting with governing MHD equations, we have to specify what kind of plasma is each medium (the moving tube and its environment).  One sees from CME parameters listed in Table~2 in Foullon et al.\ (\cite{foullon2013}) that the plasma beta inside the flux tube might be equal to $1.5 \pm 1.01$, while that of the coronal plasma is $0.21 \pm 0.05$.  Hence, we can consider the ejecta as an incompressible medium and treat its environment as a cool plasma ($\beta_{\rm e} = 0$).

The plasma motion inside the twisted flux tube is governed by the set of linearised MHD equations for an ideal incompressible plasma:
\begin{eqnarray}
\label{eq:moment}
	\rho_0 \frac{\partial}{\partial t}\vec{v}_1 + \rho_0 \left(
        \vec{v}_0 \cdot \nabla \right) \vec{v}_1 + \nabla \left( p_1
        + \frac{\vec{B}_0 \cdot \vec{B}_1}{\mu} \right) \nonumber \\
        \nonumber \\
        {}-\frac{1}{\mu} ( \vec{B}_0 \cdot \nabla ) \vec{B}_1 -
        \frac{1}{\mu} ( \vec{B}_1 \cdot \nabla ) \vec{B}_0 = 0,
\end{eqnarray}
\begin{equation}
\label{eq:induct}
	\frac{\partial}{\partial t}\vec{B}_1 - \nabla \times \left( \vec{v}_1
    \times \vec{B}_0 \right) - \nabla \times \left( \vec{v}_0 \times \vec{B}_1
    \right) = 0,	
\end{equation}
\begin{equation}
\label{eq:divv}
	\nabla \cdot \vec{v}_1 = 0,	
\end{equation}
\begin{equation}
\label{eq:divb}
	\nabla \cdot \vec{B}_1 = 0.
\vspace{3mm}
\end{equation}
Here, the index `$0$' denotes equilibrium values of the fluid velocity and the medium magnetic field, and the index `$1$' their perturbations.  Below, the sum $p_1 + \vec{B}_0 \cdot \vec{B}_1/\mu$ in Eq.~(\ref{eq:moment}) will be replaced by $p_{\rm 1 tot}$, which represents the total pressure perturbation.

As the unperturbed parameters depend on the $r$ coordinate only, the perturbations can be Fourier analyzed with $\exp \left[ \mathrm{i} \left( -\omega t + m \varphi + k_z z \right) \right]$.  Bearing in mind that in cylindrical coordinates the
nabla operator has the form
\[
	\nabla \equiv \frac{\partial}{\partial r}\hat{r} + \frac{1}{r}
        \frac{\partial}{\partial \phi}\hat{\varphi} + \frac{\partial}{\partial
        z}\hat{z},
\]
from the above set of equations one can obtain a second-order differential equation for the total pressure perturbation $p_{\rm 1 tot}$
\begin{equation}
\label{eq:diffeq}
	\left[ \frac{\mathrm{d}^2}{\mathrm{d}r^2} + \frac{1}{r}
        \frac{\mathrm{d}}{\mathrm{d} r} - \left( \kappa_{\rm i}^2 + \frac{m^2}{r^2}
        \right) \right] p_{\rm 1 tot} = 0,
\end{equation}
as well as an expression for the radial component $v_{1r}$ of the fluid
velocity perturbation $\vec{v}_1$ in terms of $p_{\rm 1 tot}$ and its
first derivative
\begin{equation}
\label{eq:v1r}
	{v}_{1r} = -\mathrm{i}\frac{1}{\rho}_{\rm i} \frac{1}{Y} \frac{\omega -
        \vec{k} \cdot \vec{v}_0}{\left( \omega - \vec{k} \cdot
        \vec{v}_0 \right)^2 - \omega_{\rm Ai}^2} \left(
        \frac{\mathrm{d}}{\mathrm{d} r} p_{\rm 1 tot} - Z \frac{m}{r}
        p_{\rm 1 tot} \right).
\end{equation}
In Eq.~(\ref{eq:diffeq}), $\kappa_{\rm i}$ is the so-called \emph{wave attenuation coefficient}, which characterises the space structure of the wave and whose squared magnitude is given by the expression
\begin{equation}
\label{eq:kappa}
	\kappa_{\rm i}^2 = k_z^2 \left(  1 - \frac{4 A^2 \omega_{\rm Ai}^2}
        {\mu \rho_{\rm i} \left[ \left( \omega - \vec{k} \cdot
        \vec{v}_0 \right)^2 - \omega_{\rm Ai}^2\right]^2} \right),
\end{equation}
where
\begin{equation}
\label{eq:alfvenfrq}
	\omega_{\rm Ai} = \frac{\vec{k}\cdot \vec{B}_{\rm i}}{\sqrt{\mu \rho}_{\rm i}} = \frac{1}{\sqrt{\mu \rho}_{\rm i}}\left( mA + k_z B_{{\rm i}z} \right)
\end{equation}
is the so-called \emph{local Alfv\'en frequency\/} (Bennett et al.\ \cite{bennett1999}).  The numerical coefficients $Z$ and $Y$ in the expression of $v_{1r}$ (see Eq.~(\ref{eq:v1r})) are respectively
\[
    Z = \frac{2A \omega_{\rm Ai}}{\sqrt{\mu \rho_{\rm i}}\left[ \left( \omega -
        \vec{k} \cdot \vec{v}_0 \right)^2 - \omega_{\rm Ai}^2\right]}
        \qquad \mbox{and} \qquad Y = 1 - Z^2.
\]

The solution to Eq.~(\ref{eq:diffeq}) bounded at the tube axis is
\begin{equation}
\label{eq:solution}
    p_{\rm tot}(r \leqslant a) = \alpha_{\rm i}I_m(\kappa_{\rm i}r),
\end{equation}
where $I_m$ is the modified Bessel function of order $m$ and $\alpha_{\rm i}$ is a constant.

For the cool environment (thermal pressure $p_{\rm e} = 0$) with straight-line magnetic field $B_{{\rm e}z} = B_{\rm e}$ and homogeneous density $\rho_{\rm e}$, from governing Eqs.~(\ref{eq:moment}), (\ref{eq:induct}), and (\ref{eq:divb}), in a similar way one obtains the same modified Bessel equation (\ref{eq:diffeq}), but $\kappa_{\rm i}^2$ is replaced by
\begin{equation}
\label{eq:kappae}
    \kappa_{\rm e}^2 = k_z^2 \left( 1 - \omega^2/\omega_{\rm Ae}^2 \right),
\end{equation}
where
\begin{equation}
\label{eq:omegaae}
    \omega_{\rm Ae} = \frac{k_z B_{{\rm e}z}}{\sqrt{\mu \rho_{\rm e}}} = k_z v_{\rm Ae},
\end{equation}
and $v_{\rm Ae} = B_{\rm e}/\sqrt{\mu \rho_{\rm e}}$ is the Alfv\'en speed in the corona.
The solution to modified Bessel equation bounded at infinity is
\begin{equation}
\label{eq:pmag}
    p_{\rm tot}(r > a) = \alpha_{\rm e}K_m(\kappa_{\rm e}r),
\end{equation}
where $K_m$ is the modified Bessel function of order $m$ and $\alpha_{\rm e}$ is a constant.

From Eq.~(\ref{eq:moment}), written down for the radial components of all perturbations, we obtain that
\[
    v_{1r} = -\mathrm{i}\frac{1}{\rho}_{\rm e} \frac{\omega}{\omega^2 - \omega_{\rm Ae}^2}\frac{\mathrm{d}}{\mathrm{d} r} p_{\rm 1 tot},
\]
or
\begin{equation}
\label{eq:v1er}
        v_{1r}(r > a) = -\mathrm{i}\frac{1}{\rho}_{\rm e} \frac{\omega}{\omega^2 - \omega_{\rm Ae}^2} \alpha_{\rm e} \kappa_{\rm e}K_m^{\prime}(\kappa_{\rm e}r),
\end{equation}
where the prime sign means a differentiation by the Bessel function argument.

To obtain the dispersion relation of MHD modes, we have to merge the solutions for $p_{\rm tot}$ and $v_{1r}$ inside and outside the tube through boundary conditions.  The boundary conditions have to ensure that the normal component of the interface perturbation
\[
    \xi_r = -\frac{v_{1r}}{\mathrm{i}\left( \omega - \vec{k} \cdot
        \vec{v}_0 \right)}
\]
remains continuous across the unperturbed tube boundary $r = a$, and also that
the total Lagrangian pressure is conserved across the perturbed boundary.  This
leads to the conditions (Bennett et al.\ \cite{bennett1999})
\begin{equation}
\label{eq:contxi}
    \xi_{{\rm i}r}|_{r=a} = \xi_{{\rm e}r}|_{r=a}
\end{equation}
and
\begin{equation}
\label{eq:contp1tot}
    \left.p_{\rm tot\,i} - \frac{B_{{\rm i}\varphi}^2}{\mu a}\xi_{{\rm i}r}\right\vert_{r=a} = p_{\rm tot\,e}|_{r=a},
\end{equation}
where total pressure perturbations $p_{\rm tot\,i}$ and $p_{\rm tot\,e}$ are given by Eqs.~(\ref{eq:solution}) and (\ref{eq:pmag}), respectively.  Applying boundary conditions (\ref{eq:contxi}) and (\ref{eq:contp1tot}) to our solutions of $p_{\rm 1 tot}$ and $v_{1r}$ (and accordingly $\xi_r$), we obtain after some algebra the dispersion relation of the normal modes propagating along a twisted magnetic tube with axial mass flow $\vec{v}_0$
\begin{eqnarray}
\label{eq:dispeq}
	\frac{\left[ \left( \omega - \vec{k}\cdot \vec{v}_0 \right)^2 -
    \omega_{\rm Ai}^2 \right]F_m(\kappa_{\rm i}a) - 2mA \omega_{\rm Ai}/\sqrt{\mu \rho_{\rm i}}}
    {\left[ \left( \omega - \vec{k}\cdot \vec{v}_0 \right)^2 -
    \omega_{\rm Ai}^2 \right]^2 - 4A^2\omega_{\rm Ai}^2/\mu \rho_{\rm i} } \nonumber \\
    \nonumber \\
    {} = \frac{P_m(\kappa_{\rm e} a)}
    {{\displaystyle \frac{\rho_{\rm e}}{\rho_{\rm i}}} \left( \omega^2 - \omega_{\rm Ae}^2
    \right) + A^2  P_m(\kappa_{\rm e} a)/\mu \rho_{\rm i}},
\end{eqnarray}
where, $\omega - \vec{k}\cdot \vec{v}_0$ is the Doppler-shifted wave frequency in the moving medium, and
\[
    F_m(\kappa_{\rm i}a) = \frac{\kappa_{\rm i}a I_m^{\prime}(\kappa_{\rm i}a)}{I_m(\kappa_{\rm i}a)} \quad \mbox{and} \quad P_m(\kappa_{\rm e}a) = \frac{\kappa_{\rm e}a K_m^{\prime}(\kappa_{\rm e}a)}{K_m(\kappa_{\rm e}a)}.
\]
This dispersion equation is similar to the dispersion equation of normal MHD modes in a twisted flux tube surrounded by incompressible plasma (Zhelyazkov \& Zaqarashvili \cite{zhelyazkov2012}) -- there, in Eq.~(13), $\kappa_{\rm e} = k_z$, and to the dispersion equation for a twisted tube with non-magnetized environment, i.e., with $\omega_{\rm Ae} = 0$ (Zaqarashvili et al.\ \cite{zaqarashvili2010}).

\section{Numerical solutions and wave dispersion diagrams}
\label{sec:dispers}
The main goal of our study is to see under which conditions the propagating along the moving flux tube MHD waves can become unstable.  To conduct such an investigation it is necessary to assume that the wave frequency $\omega$ is a complex quantity, i.e., $\omega \to \omega + \mathrm{i}\gamma$, where $\gamma$ is the instability growth rate, while the longitudinal wave number $k_z$ is a real variable in the wave dispersion relation.  Since the occurrence of the expected KH instability is determined primarily by the jet velocity, in looking for a critical/threshold value of it, we will gradually change its magnitude from zero to that critical value (and beyond).  Thus, we have to solve dispersion relations in complex variables obtaining the real and imaginary parts of the wave frequency, or as it is normally accepted, of the wave phase velocity $v_{\rm ph} = \omega/k_z$, as functions of $k_z$ at various values of the velocity shear between the surge and its environment, $v_0$.

We focus our study first on the propagation of the kink mode, i.e., for $m = 1$.  It is obvious that Eq.~(\ref{eq:dispeq}) can be
solved only numerically.  The necessary step is to define the input parameters which characterise the moving twisted magnetic flux tube, and also to normalise all variables in the dispersion equation.  The density contrast between the tube and its environment is characterised by the parameter $\eta = \rho_{\rm e}/\rho_{\rm i}$, and the twisted magnetic field by the ratio of the two magnetic field components, $B_{{\rm i}\varphi}$ and $B_{{\rm i}z}$, evaluated at the inner boundary of the tube, $r = a$, i.e., via $\varepsilon = B_{{\rm i}\varphi}/B_{{\rm i}z}$, where $B_{{\rm i}\varphi} = Aa$.  As usual, we normalise the velocities to the Alfv\'en speed $v_{\rm Ai} = B_{{\rm i}z}/(\mu \rho_{\rm i})^{1/2}$.  Thus, we introduce the dimensionless wave phase velocity $v_{\rm ph}/v_{\rm Ai}$ and the \emph{Alfv\'en Mach number\/} $M_{\rm A} = v_0/v_{\rm Ai}$, the latter characterising the axial motion of the tube.  The wavelength, $\lambda = 2\pi/k_z$, is normalised to the tube radius $a$ which implies that the dimensionless wave number is $k_z a$.  The natural way of normalising the local Alfv\'en frequency $\omega_{\rm Ai}$ is to multiply it by the tube radius, $a$, and after that divide by the Alfv\'en speed $v_{\rm Ai}$, i.e.,
\[
    a \omega_{\rm Ai} = \frac{B_{{\rm i}z}}{\sqrt{\mu \rho_{\rm i}}}\left( m\frac{B_{{\rm i}\varphi}}{B_{{\rm i}z}} + k_z a \right) \quad \mbox{or} \quad \frac{a \omega_{\rm Ai}}{v_{\rm Ai}} = m \varepsilon + k_z a,
\]
where $B_{{\rm i}\varphi}/B_{{\rm i}z} \equiv Aa/B_{{\rm i}z} = \varepsilon$ is the above defined twist parameter.  We note that the normalisation of Alfv\'en frequency outside the jet, $\omega_{\rm Ae}$, requires (see Eq.~(\ref{eq:omegaae})) except the tube radius, $a$, and the density contrast, $\eta$, the ratio of the two axial magnetic fields, $b = B_{\rm e}/B_{{\rm i}z}$.

Before starting the numerical job, we have to specify the values of the input parameters.  Our choice for the density contrast is $\eta = 0.88$, which corresponds to electron densities $n_{\rm i} = 8.7 \times 10^8$~cm$^{-3}$ and $n_{\rm e} = 7.67 \times 10^8$~cm$^{-3}$, respectively.  With $\beta_{\rm i} = 1.5$ and $\beta_{\rm e} = 0$, the ratio of axial magnetic fields is $b = 1.58$.  If we fix the Alfv\'en speed in the environment to be $v_{\rm Ae} \cong 787$~km\,s$^{-1}$ (i.e., the value corresponding to $n_{\rm e} = 7.67 \times 10^8$~cm$^{-3}$ and $B_{\rm e} = 10$~G), the total pressure balance equation at $\eta = 0.88$ requires a sound speed inside the jet $c_{\rm si} \cong 523$~km\,s$^{-1}$ and Alfv\'en speed $v_{\rm Ai} \cong 467$~km\,s$^{-1}$ (more exactly, $467.44$~km\,s$^{-1}$), which corresponds to a magnetic field in the flux tube $B_{{\rm i}z} = 6.32$~G.  Following Ruderman (\cite{ruderman2007}), to satisfy the Shafranov-Kruskal stability criterion for a kink instability we assume that the azimuthal component of the magnetic field $\vec{B}_{\rm i}$ is smaller than its axial component, i.e., we will choose our twist parameter $\varepsilon$ to be always less than $1$.  In particular, we will study the dispersion diagrams of kink, $m = 1$ and $m = -3$ MHD modes and their growth rates (when the modes are unstable) for three fixed values of $\varepsilon$: $0.025$, $0.1$, and $0.2$.

\begin{figure}[ht]
\centering
  \begin{minipage}{\columnwidth}
  \centering
    \includegraphics[width=7.5cm]{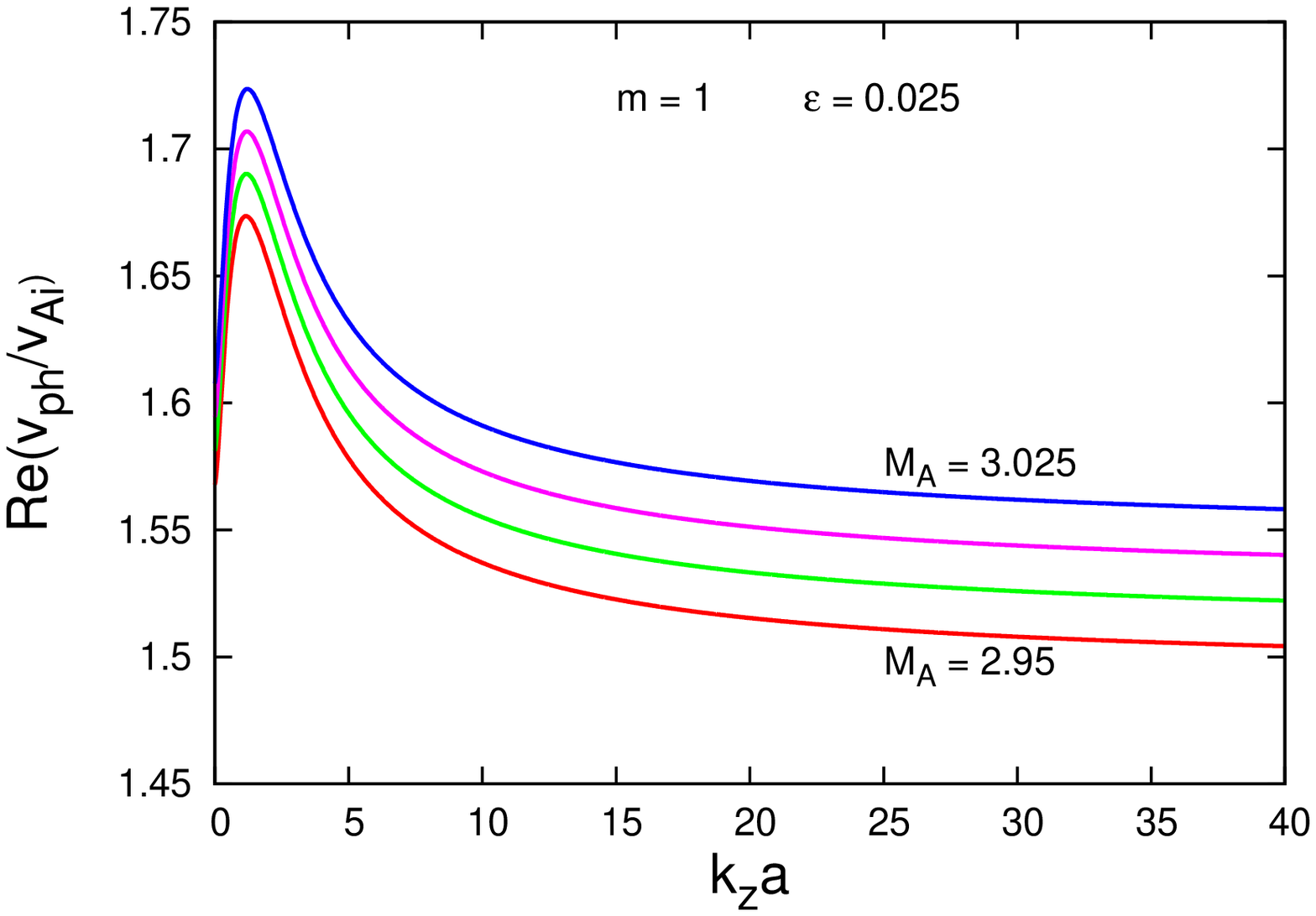} \\
\vspace{0mm}
    \includegraphics[width=7.5cm]{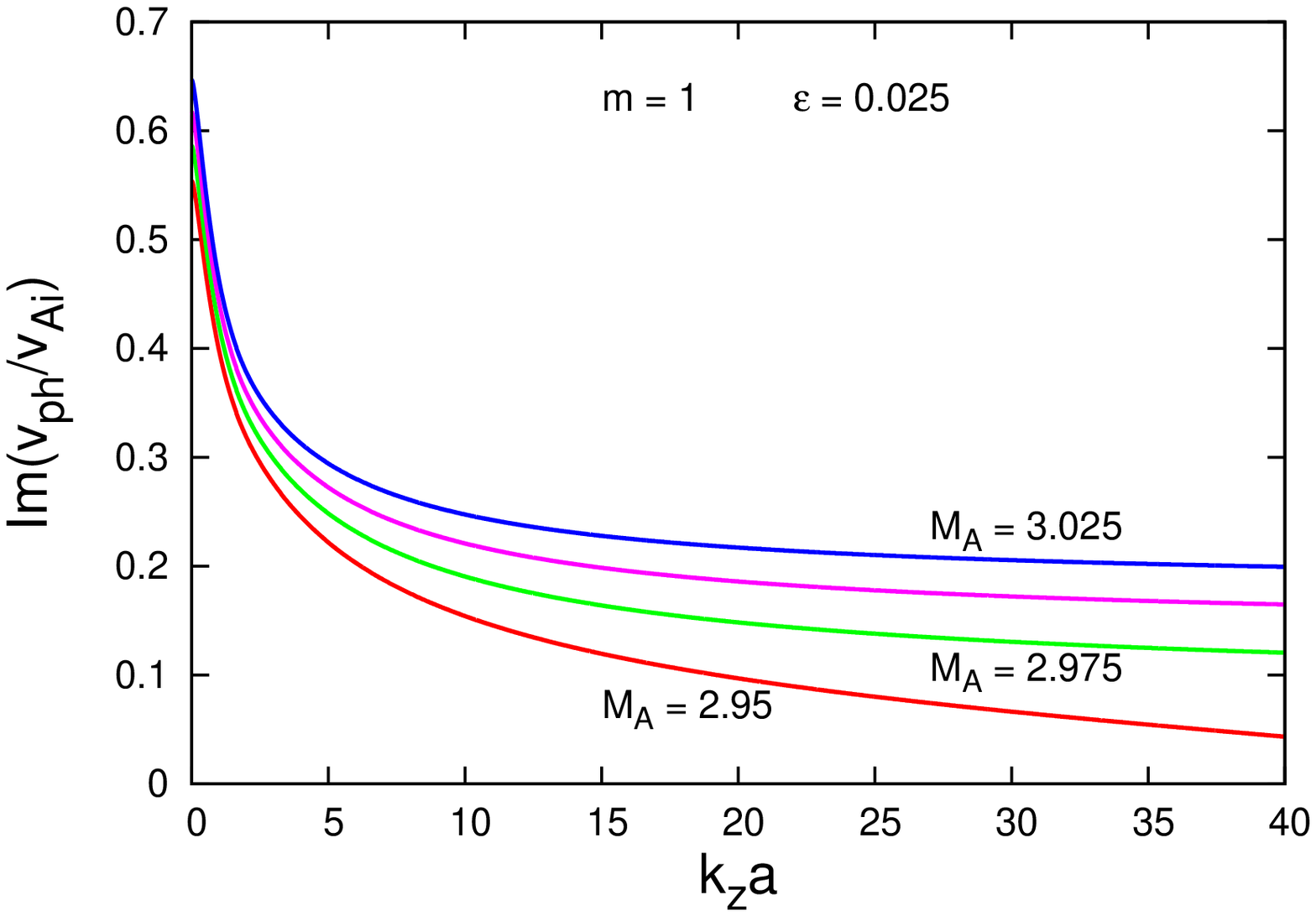}
  \end{minipage}
  \caption{(\emph{Top panel}) Dispersion curves of unstable kink ($m = 1$) mode propagating on a moving twisted magnetic flux tube with twist parameter $\varepsilon = 0.025$ at $\eta = 0.88$, $b = 1.58$ and Alfv\'en Mach numbers equal to $2.95$, $2.975$, $3$, and $3.025$, respectively. (\emph{Bottom panel}) Growth rates of the unstable kink mode.  Note that the marginal red curve reaches values smaller than $0.001$ beyond $k_z a = 53$.}
  \label{fig:fig4}
\end{figure}
\begin{figure}[ht]
\centering
  \begin{minipage}{\columnwidth}
  \centering
    \includegraphics[width=7.5cm]{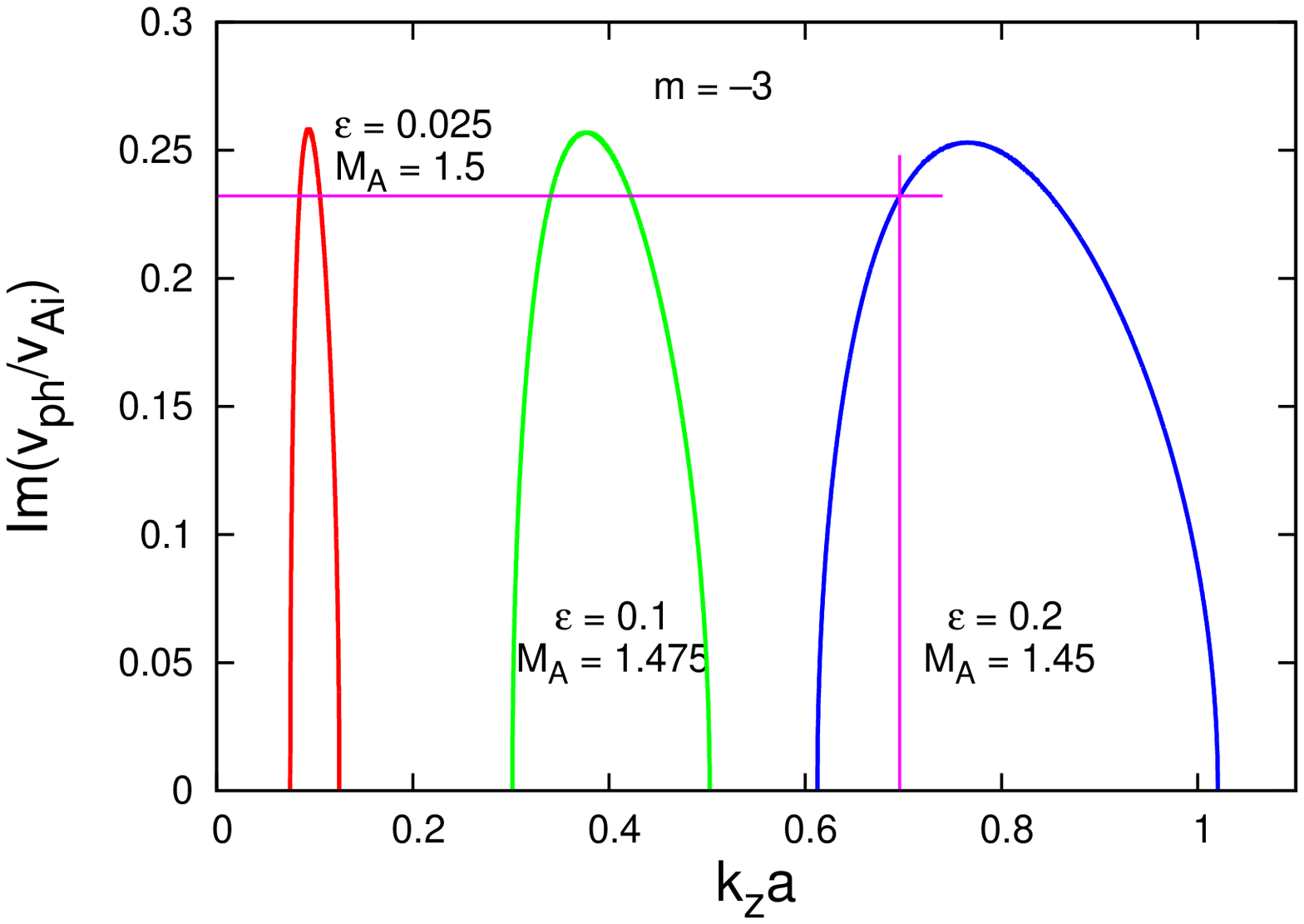} \\
\vspace{0mm}
    \includegraphics[width=7.5cm]{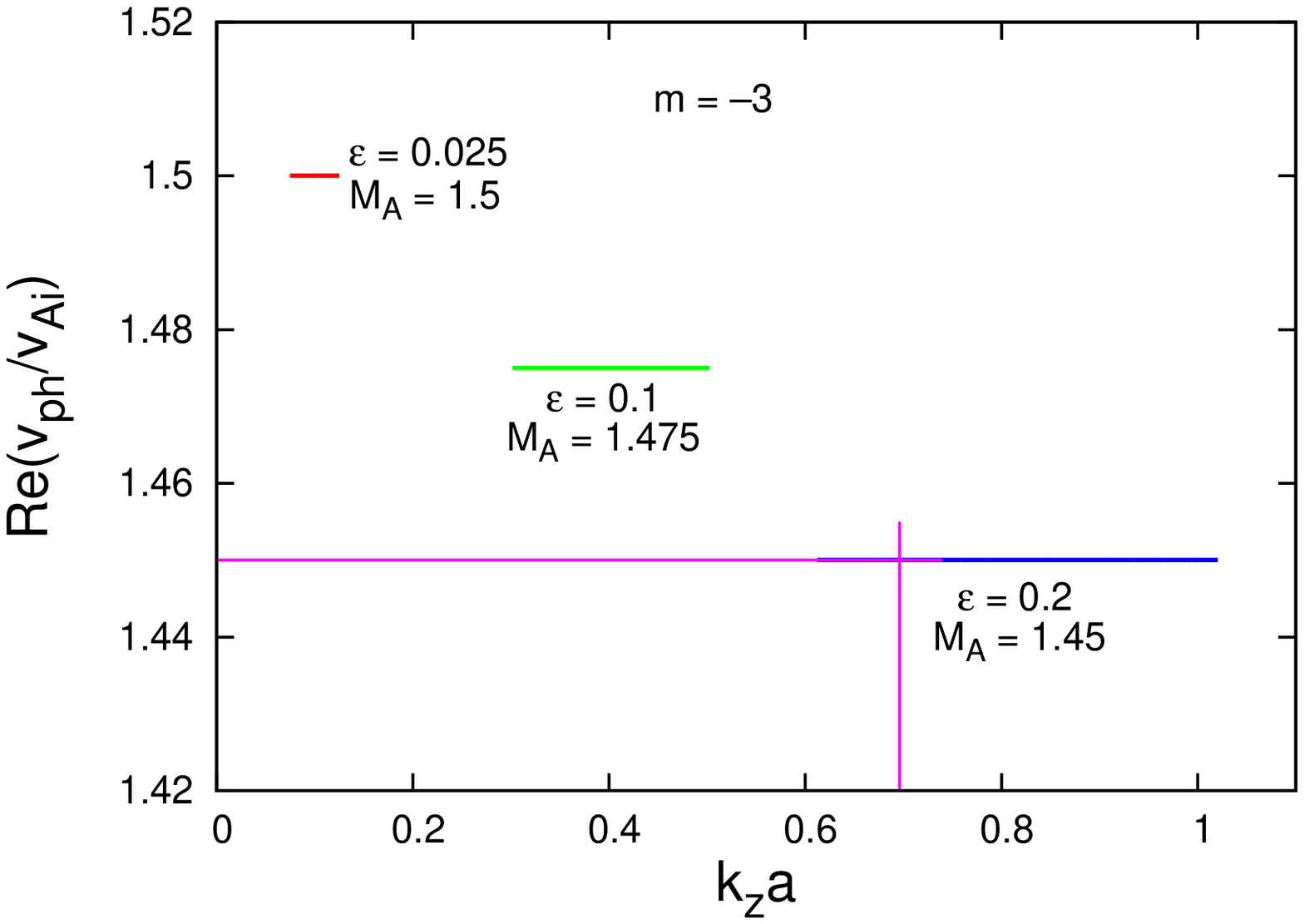}
  \end{minipage}
  \caption{(\emph{Top panel}) Growth rates of unstable $m = -3$ MHD mode in three instability windows.   Adequate numerical values consistent with the observational data one obtains at $k_z a = 0.696245$, located in the third instability window, for which value of $k_z a$ the normalized mode growth rate is equal to $0.232117$. (\emph{Bottom panel}) Dispersion curves of unstable $m = -3$ MHD mode for three values of the twist parameter $\varepsilon = 0.025$, $0.1$, and $0.2$. The normalized phase velocity at $k_z a = 0.696245$ is equal to $1.45$.}
  \label{fig:fig5}
\end{figure}
With these input parameters ($M_{\rm A}$ will be varied from zero (no flow) to reasonable values during computations), the solutions to Eq.~(\ref{eq:dispeq}) yield the dependence of the normalized wave phase velocity $\omega/k_z v_{\rm Ai}$ on $k_z a$.
In Fig.~\ref{fig:fig4} we display the dispersion curves of the kink mode at $\varepsilon = 0.025$ for various values of $M_{\rm A}$ -- the red curve is in fact a \emph{marginal\/} dispersion curve corresponding to the critical $M_{\rm A}^{\rm cr} = 2.95$.  The critical jet speed, $M_{\rm A}^{\rm cr}v_{\rm Ai} \cong 1380$~km\,s$^{-1}$, being in principal accessible for CMEs, is, in fact, more than $2$ times higher than the registered by Foullon et al.\ (\cite{foullon2013}) threshold speed of $680$~km\,s$^{-1}$ for observing KH instability.  For the next higher values of $\varepsilon$ one obtains threshold Alfv\'en Mach numbers around $3$.  Hence, detected KH instability cannot be associated with the kink mode.  But situation distinctly changes for the $m = -3$ MHD mode.  As seen from Fig.~\ref{fig:fig5}, one can observe the appearance of three instability windows on the $k_z a$-axis.  The width of each instability window depends upon the value of twist parameter $\varepsilon$ -- the narrowest window corresponds to $\varepsilon = 0.025$, and the widest to $\varepsilon = 0.2$.  It is worth noticing that the phase velocities of unstable $m = -3$ MHD modes coincides with the magnetic flux tube speeds (one sees in the bottom panel of Fig.~\ref{fig:fig5} that the normalized wave velocity on given dispersion curve is equal to its label $\mathsf{M_{\sf A}}$).  Therefore, unstable perturbations are frozen in the flow and consequently they are vortices rather than waves.  It has a firm physical ground as KH instability in hydrodynamics deals with unstable vortices.  All critical Alfv\'en Mach numbers yield acceptable threshold speeds of the ejecta which ensure the occurrence of KH instability -- these speeds are equal to $701$~km\,s$^{-1}$, $689$~km\,s$^{-1}$, and $678$~km\,s$^{-1}$, respectively, in very good agreement with the speed of $680$~km\,s$^{-1}$ found by Foullon et al.\ (\cite{foullon2013}).  The observationally detected KH instability wavelength $\lambda_{\rm KH} = 18.5$~Mm and ejecta width $\Delta L = 4.1$~Mm define the corresponding instability dimensionless wave number, $k_z a = \pi \Delta L/\lambda$, to be equal to $0.696245$.  As seen from Fig.~\ref{fig:fig5}, that $k_z a = 0.696245$ lies in the third instability window and accordingly determines a value of the dimensionless growth rate Im$(v_{\rm ph}/v_{\rm Ai}) = 0.232117$ (see the top panel of Fig.~\ref{fig:fig5}), which implies a computed wave growth rate $\gamma_{\rm KH} = 0.037$~s$^{-1}$, being in good agreement with the deduced from observations $\gamma_{\rm KH} = 0.033$~s$^{-1}$.  We note also that the estimated from Fig.~\ref{fig:fig5} (bottom panel) wave phase velocity of $678$~km\,s$^{-1}$ is rather close to the speed of the $131$~\AA~CME leading edge equal to $687$~km\,s$^{-1}$.  It is worth pointing out that the position of a given instability window, at fixed input parameters $\eta$ and $b$, is determined chiefly by the magnetic field twist in the moving flux tube.  This circumstance allows us by slightly shifting the third instability window to the right, to tune the vertical purple line (see the top panel of Fig.~\ref{fig:fig5}) to cross the growth rate curve at a value, which would yield $\gamma_{\rm KH} = 0.033$~s$^{-1}$.  The necessary shift of only $0.023625$ can be achieved by taking the magnetic field twist parameter, $\varepsilon$, to be equal to $0.20739$ -- in that case the normalised Im$(v_{\rm ph}/v_{\rm Ai}) = 0.207999$ gives the registered KH instability growth rate of $0.033$~s$^{-1}$.  (That very small instability window shift does not change noticeably the critical ejecta speed.)  In this way, we demonstrate the flexibility of our model allowing the derivation of numerical KH instability characteristics in very good agreement with observational data.

\section{Discussion and conclusion}
\label{sec:concl}
In this paper, as we have shown, the imaged by Foullon et al.\ (\cite{foullon2013}) KH vortices on 2010 November 3 CME can be explained as KH instability of the $m = -3$ harmonic in a twisted flux tube moving in external cool magnetized plasma embedded in homogeneous untwisted magnetic field.  We have assumed the wave vector $\vec{k}$ aligned with the $\vec{v}_0 = \vec{V}_{\rm i} - \vec{V}_{\rm e}$ vector.  We would like to point out that the results of the numerical modelling crucially depend on the input parameters.  Any small change in the density contrast, $\eta$, or the magnetic fields ratio, $b$, can dramatically change the picture.  In our case, the input parameters for solving the MHD mode dispersion relation in complex variables (the mode frequency $\omega$ and, respectively, the mode phase velocity $v_{\rm ph} = \omega/k_z$, were considered as complex quantities) were chosen to be consistent with the plasma and magnetic field parameters listed in Table~2 in Foullon et al.\ (\cite{foullon2013}).  It was found that with a twist parameter of the background magnetic filed $\vec{B}_{\rm i}$, $\varepsilon = 0.2$, the critical jet's speed is $v_0^{\rm cr} = 678$~km\,s$^{-1}$, and at wavelength of the unstable $m = -3$ mode $\lambda_{\rm KH} = 18.5$~Mm and ejecta width $\Delta L = 4.1$~Mm, its growth rate is $\gamma_{\rm KH} = 0.037$~s$^{-1}$.  These values of $v_0^{\rm cr}$ and $\gamma_{\rm KH}$ are in a good agreement with the data listed in Table~3 of Foullon et al.\ (\cite{foullon2013}).  We have also shown that the numerically obtained instability growth rate can be a little reduced to coincide with observational one of $0.033$~s$^{-1}$ through slightly shifting to the right the appropriate instability window -- this can be done by performing the calculations with a new value of the magnetic field twist parameter $\varepsilon$, equal to $0.20739$.  Thus, our model is flexible enough to allow us numerically get KH instability characteristics very close to the observed ones.  The two ``cross points'' in Fig.~\ref{fig:fig5} can be considered as a `computational portrait' of the imaged on the 2010 November 3 coronal mass ejecta KH instability.  Critical ejecta speed and KH instability growth rate values, in good agreement with those derived by Foullon et al.\ (\cite{foullon2013}), can also be obtained by exploring the flute-like, $m = -2$, MHD mode (Zhelyazkov \& Chandra \cite{zhelyazkov2014}).  In that case, a better agreement with the observational data one achieves at $\varepsilon = 0.188$ -- the computed values of $v_0^{\rm cr}$ and $\gamma_{\rm KH}$ are exactly the same as those for the $m = -3$ mode.

It is necessary to stress that each CME is a unique event and its successful modelling requires a full set of observational data for the plasma densities, magnetic fields, and temperatures of both media along with the detected ejecta speeds.  Concerning values of the flux tube speeds at which the instability starts, they can vary from a few kilometers per second, $6$--$10$~km\,s$^{-1}$, as observed by Ofman \& Thompson (\cite{ofman2011}), through $310 \pm 20$~km\,s$^{-1}$ of M\"{o}stl et al.\ (\cite{meostl2013}), to $680 \pm 92$~km\,s$^{-1}$ deduced from Foullon et al.\ (\cite{foullon2013}).  It is curious to see whether the $13$ fast flareless CMEs (with velocities of $1000$~km\,s$^{-1}$ and higher) observed from 1998 January 3 to 2005 January 4 (see Table~1 in Song et al.\ (\cite{song2013})), are subject to the KH instability.

In spite of the fact that KH instability characteristics of both $m = -2$ and $m = -3$ MHD modes are in a good agreement with observational data, only the instability of the $m = -3$ harmonic may explain why the KH vortices are seen only at one side of rising CME (Foullon et al.\ (\cite{foullon2011,foullon2013}), see also Fig.~\ref{fig:fig2} in the present paper).  This harmonic yields that the unstable vortices have $3$ maxima around the magnetic tube with a $360/3 = 120$ degree interval.  Therefore, if one maximum is located in the plane perpendicular to the line of sight (as it is clearly seen by observations), then one cannot detect two other maxima in imaging observations as they will be largely directed along the line of sight.

Finally we would like to comment on whether there is a required fields orientations $\phi = \widehat{(\vec{k}, \vec{B})}$.  Analysing a flat (semi-infinite) geometry in their study, Foullon et al.\ (\cite{foullon2013}) conclude that it is most likely to observe a quasi-perpendicular to the magnetic field $\vec{B}_{\rm e}$ wave propagation.  But this restriction on the ejecta magnetic field tilt angle drops out in our numerical investigation because the inequality $|V_{\rm i} - V_{\rm e}| \geqslant \sqrt{2}v_{\rm Ai} = 661$~km\,s$^{-1}$, required for wave parallel propagation, is satisfied.  Thus, the adopted magnetic flux rope nature of a CME and its (ejecta) consideration as a moving twisted magnetic flux tube allow us to explain the emerging instability as a manifestation of the KH instability of a suitable MHD mode.

\begin{acknowledgements}
The work of two of us, IZh and RC, was supported by the Bulgarian Science Fund and the Department of Science \& Technology, Government of India Fund under Indo-Bulgarian bilateral project CSTC/INDIA 01/7, /Int/Bulgaria/P-2/12.  The work of TZ was supported by the European FP7-PEOPLE-2010-IRSES-269299 project-SOLSPANET, by the Austrian Fonds zur F\"{o}rderung der Wissenschaftlichen Forschung (project P26181-N27), by Shota Rustaveli National Science Foundation grant DI/14/6-310/12 and by Austrian Scientific-technology collaboration (WTZ) grant IN 10/2013.  We would like to thank the referee for the very useful comments and suggestions which help us improve the quality of our paper, and are indebted to Dr.~Snezhana Yordanova for drawing one figure.
\end{acknowledgements}

\end{document}